\documentclass[reprint,aps,prc,amsmath,amssymb,showpacs,showkeys,twocolumns,floatfix,diagbox]{revtex4-1}
\usepackage{graphicx}
\usepackage{dcolumn}
\usepackage{bm}
\usepackage{longtable}
\usepackage{multirow}
\usepackage[flushleft]{threeparttable}
\begin{document}

\preprint{APS/123-QED}

\title{Critical view of nonlocal nuclear potentials in alpha decay}

\title{Critical view of nonlocal nuclear potentials in alpha decay}
\author{J. E. Perez Velasquez}
\email{je.perez43@uniandes.edu.co}
\affiliation{Departamento de Fisica, Universidad de los Andes,
Cra.1E No.18A-10, Bogot\'a, Colombia}
\author{N. G. Kelkar}
\email{nkelkar@uniandes.edu.co}
\affiliation{Departamento de Fisica, Universidad de los Andes,
Cra.1E No.18A-10, Bogot\'a, Colombia}
\author{N. J. Upadhyay}
\email{neelam.upadhyay@cbs.ac.in}
\affiliation{School of Physical Sciences, UM-DAE Centre for 
Excellence in Basic Sciences, Vidyanagari, Mumbai-400098, India}
\date{\today}

\begin{abstract}
Different models for the nonlocal description of the nuclear interaction 
are compared through a study of their effects on the half-lives 
of radioactive nuclei decaying by the emission of alpha particles. 
The half-lives are evaluated by considering a  
pre-formed alpha particle ($^4$He nucleus) which tunnels through the 
Coulomb barrier generated by its interaction with the daughter nucleus. 
An effective potential obtained from a density dependent 
double folding strong potential 
between the alpha and the daughter nucleus within the nonlocal framework 
is found to decrease the 
half-lives as compared to those in the absence of nonlocalities.
Whereas the percentage decrease within the older Perey-Buck and 
S\~ao Paulo models ranges between 20 to 40\% for medium to heavy nuclei,  
a recently proposed effective potential leads to
a decrease of only 2 - 4 \%. In view of these results, we provide a closer
examination of the approximations used in deriving the local equivalent 
potentials and propose that apart from the scattering data, 
the alpha decay half-lives could be used as a
complementary tool for constraining the nonlocality models.    
\end{abstract}



\maketitle

\section{Introduction}
It is not often that revisiting an old and well studied subject reveals 
new findings. However, one does find examples of 
experimental as well theoretical investigations,  
which, either with more refined tools or alternative theoretical approaches 
attempt to probe into supposedly established methods to bring new 
results and solutions. The cosmological lithium problem is a
recent example of such a situation where conventional methods overestimated 
the $^7$Li abundance but the introduction of Tsallis statistics within 
these methods solved the problem \cite{bertulani}.  Another recent example is that 
of pinning down the D-state probability in the deuteron (a topic which has been a
classic problem of nuclear physics) using modern precise measurements of the 
Lamb shift in the muonic deuterium atom \cite{mePLB}. 
In the context of the present work, we note that the models for the 
strong nuclear interaction within the nonlocal framework,  
have been studied since decades with the pioneering works in Refs. \cite{bethe, 
frahn,vinhmau}. Perey and Buck \cite{pereybuck} studied scattering 
using the nonlocal framework and introduced a local equivalent potential. 
These works were followed up by several others which 
were able to reproduce the scattering data quite well \cite{balantekin,
chamon1}. However, revisiting the nonlocality within a novel approach 
to the same problem, Refs. \cite{neelam1,neelam2} revealed some interesting 
features. To list a few, the framework is flexible to use arbitrary nonlocal 
potentials, is not sensitive to the choice of the nonlocal form factor \cite{neelam2}  
and the effective potential has a different behaviour in 
coordinate ($r$) space for $r \to 0$ as compared to the local equivalent potentials in Refs. \cite{pereybuck} and \cite{chamon1}.

Another well-established method in nuclear physics is the treatment of 
alpha decay as a tunneling problem for the calculation of half-lives of 
nuclei with a density dependent double folding 
(DF) potential \cite{weWKB,weEPL,weMagic}. This method is quite successful 
in reproducing the half-lives of a range of medium and heavy nuclei 
\cite{nirenreview,xurenold}. However, the effects of nonlocality have not 
been studied within this model. 
In the present work, starting from the DF potential between the alpha 
($^4$He nucleus) and the daughter which exist as a pre-formed cluster 
inside the decaying parent nucleus, we obtain effective 
potentials in three different models and study their effects on the 
alpha decay half-lives. Though the general finding from all models 
is a decrease in the half-lives due to nonlocality, the percentage decrease 
using the model of Ref \cite{neelam1} is significantly smaller than that 
with \cite{pereybuck,chamon1}.  

The article is organized as follows: 
in Section \ref{alphadecay} we present the 
density dependent double folding model used to evaluate the potential 
between the alpha and the daughter nucleus followed by the
formalism for the evaluation of alpha decay half-lives within a semiclassical 
approach to the tunneling problem.
In Section \ref{nonlocalpots}, 
the concept of nonlocality and the three models used in the present 
work are briefly introduced. Two of the models \cite{pereybuck,chamon1} are 
found to differ 
significantly from the model of Ref. \cite{neelam1} at small distances.    
Section \ref{smalldistance} explains the reason behind the discrepancies in the 
behaviour of the local equivalent potentials for $r\to 0$.  
Section \ref{iterations} briefly describes the iterative scheme used for the 
determination of the scattering wave function and a possible extension to 
the case of decaying states.  
In Section \ref{results}, we present the results 
and discuss them. Finally, in Section \ref{conclude} we summarize our 
findings. 
Since the nonlocality has no
particular importance for elastic scattering and 
is expected to affect cross sections for reaction 
processes such as stripping and inelastic scattering \cite{glenden}, 
we propose that the data on the half-lives 
of radioactive nuclei could be used as a complementary tool 
in addition to the scattering data which are 
generally used to restrict the nonlocality models. 

\section{Formalism for alpha decay} \label{alphadecay}
The objective of the present work is to examine the differences between the 
existing models to evaluate effective
potentials in the nonlocal framework through their effects on the 
half-lives of radioactive nuclei which decay by alpha particle emission.
In order to evaluate these effective 
potentials, we shall use the density dependent double-folding alpha-nucleus potential 
\cite{weWKB} which is often used in calculations of alpha decay \cite{nirenreview}. 
We assume the existence of a pre-formed alpha 
inside the parent and consider the alpha decay to be a tunneling problem of the 
alpha through the Coulomb barrier created by its interaction with the daughter nucleus.
Typically, one considers the tunneling of the $\alpha$
through an $r$-space potential of the form,
\begin{equation}\label{pot1}
V(r)\, =\, 
V_n(r)\,+\,V_C(r)\,+\,{\hbar^2\,(l\,+\,1/2)^2 \over \mu\,r^2}\, ,
\end{equation}
where $V_n(r)$ and $V_C(r)$ are the nuclear and Coulomb parts of the
$\alpha$-nucleus (daughter) potential, $r$ the distance between
the centres of mass of the daughter nucleus and alpha and $\mu$
their reduced mass.  
The last term
represents the Langer modified centrifugal barrier \cite{langer}. 
The width of the radioactive nucleus or its half-life which is related to 
it, is 
evaluated with a semiclassical JWKB approach \cite{froeman}. 
With the JWKB being valid for one-dimensional problems, the above modification 
of the centrifugal barrier from
$l(l+1) \,\rightarrow \,(l+1/2)^2$ is essential to ensure the correct
behaviour of the JWKB radial wave function near the origin as well as
the validity of the connection formulas used \cite{morehead}.

Since the aim of the present work 
is to compare the nonlocal effects in different models, we shall 
restrict here to the simpler situations of alpha decay of spherical nuclei in 
the $s$-wave. 
\subsection{The alpha nucleus double-folding potential}\label{doublefold}
The input to the double-folding 
model is a realistic nucleon-nucleon interaction as given in \cite{satchler}.
The folded nuclear potential is written as,
\begin{equation}\label{potnucl}
V_n(r)=\lambda \int d{\bf r}_1\, d{\bf r}_2\,
\rho_{\alpha}({\bf r}_1) \, \rho_d({\bf r}_2)\,v({\bf r}_{12}={\bf r}
+{\bf r}_2-{\bf r}_1,\,E)
\end{equation}
where $\rho_{\alpha}$ and $\rho_{d}$ are the densities of the alpha and the
daughter nucleus in a decay, 
$|\bf{r}_{12}|$ is the distance between a nucleon in the alpha 
and a nucleon in the daughter nucleus and, $v(\bf{r}_{12},E)$, is 
the M3Y nucleon-nucleon (NN) interaction \cite{satchler} 
given as,
\begin{eqnarray}
\nonumber
v({\bf r}_{12},E) &=& 7999\,{exp(-4\,|{\bf r}_{12}|) 
\over 4\,|{\bf r}_{12}|} 
\, -\, 2134\, {exp(-2.5\,|{\bf r}_{12}|) \over 2.5\,|{\bf r}_{12}|}
\\
\label{nnfree}
&&\,+\,J_{00}\,\delta({\bf r}_{12})\,,
\\ 
\nonumber
{\rm with\,\,\,\,}
J_{00}&=&\,-276\,(1\,-\,0.005\,E_{\alpha}/A_{\alpha})\,, 
\end{eqnarray}
where the last term is the so-called ``knock-on exchange" term and is usually not
included in the calculation of nonlocal nuclear potentials \cite{chamon1,neelam1}. 

The alpha particle density is given using a standard Gaussian form \cite{satchler}, 
namely, 
\begin{equation}\label{alphadens}
\rho_{\alpha} (r)\,=\,0.4229\,exp(-0.7024\,r^2)
\end{equation}
and the daughter nucleus density is taken to be,
\begin{equation}\label{daughter}
\rho_d(r)\,=\, {\rho_0 \over 1\,+\,exp({r - c \over a})}
\end{equation}
where $\rho_0$ is obtained by normalizing $\rho_d(r)$ to the number of 
nucleons $A_d$ and the constants are given
as $c\,=\,1.07\,A_d^{1/3}$ fm and $a\,=\,0.54$ fm \cite{paras}. 
Equation (\ref{potnucl}) 
involves a six dimensional integral. However, the numerical evaluation becomes
simpler if one works in momentum space as shown in \cite{satchler}. 
The constant $\lambda$ appearing in Eq. (\ref{potnucl}) for the 
nuclear potential $V_n(r)$ (which is a part of the total potential 
$V(r)$ in Eq. (\ref{pot1})), 
is determined by imposing the Bohr-Sommerfeld 
quantization condition:
\begin{equation}\label{bohrsomer}
\int_{r_1}^{r_2} \,\, k(r)\,dr\,=\,(n\,+\,1/2)\,\pi
\end{equation}
where $k(r) \,=\, \sqrt{{2\mu \over \hbar^2}\,(|V(r)\,-\,E)|}$,  
$n$ is the
number of nodes of the quasibound wave function of $\alpha$-nucleus
relative motion and $r_1$ and $r_2$ which are solutions of $V(r) \,=\,E$, are
the classical turning points. 
This condition is a requisite for the correct use 
of the JWKB approximation \cite{weWKB}. 
The number of nodes are re-expressed as $n\, = \,(G\,-\,l)\,/2$, where 
$G$ is a global quantum number obtained from fits to data \cite{buck} and
$l$ is the orbital angular momentum quantum number. We choose the values of 
$G$ as 18 for N $<$ 82, 20 for 82 $<$ N $\le$ 126 and 22 for 
N $>$ 126 as recommended in \cite{buck}. 

The Coulomb potential, $V_C(r)$, 
is obtained by using a similar double folding procedure
with the matter densities of the alpha and the daughter replaced by their
respective charge density distributions $\rho^C_{\alpha}$ 
and $\rho^C_{d}$. Thus,  
\begin{equation}\label{potcol}
V_C(r)\,=\,\int\,d{\bf r}_1\,d{\bf r}_2\,
\rho^C_{\alpha}({\bf r}_1) \, \rho^C_{d}({\bf r}_2)\,{e^2 \over |{\bf r}_{12}|}
\,.
\end{equation}
The charge distributions are taken to have a similar form as the matter 
distributions, except for the fact that they are normalized to the number of
protons in the alpha and the daughter. 
\subsection{Semi-classical approach for half-lives}\label{halflife}
Considering the alpha decay to be a tunneling problem, the semi-classical 
expression for the decay width as obtained from different approaches agrees and is 
given by \cite{weWKB}:
\begin{equation}\label{gurwidth}
\Gamma(E)\,= \, P_{\alpha}\,{\hbar^2 \over 2 \,\mu}\,\,\biggl[\,\int_{r_1}^{r_2}\,
{dr \over k(r)}\,\biggr]^{-1}\,e^{-2\int_{r_2}^{r_3}\,k(r)\,dr}
\end{equation}
where, $k(r) \,=\, \sqrt{{2\mu \over \hbar^2}\,(|V(r)\,-\,E|)}$. 
The energy $E$ is taken to be the same as the $Q$ value for a given alpha 
decay. The factor in
front of the exponential arises from the normalization of the bound state
wave function in the region between the turning points $r_1$ and $r_2$. 
The alpha decay half-life of a nucleus is evaluated as 
\begin{equation}\label{halflifeeq} 
\tau_{1/2}^{theory} \, =\, {\hbar \, {\rm ln \,2} \over \Gamma }\, .
\end{equation}

The factor $P_{\alpha}$ in Eq. (\ref{gurwidth}) 
takes into account the probability for the existence of a pre-formed cluster of the 
alpha and the nucleus. 
This factor, in principle, can be expressed 
as an overlap between the wave function of the parent nucleus 
and the decaying-state wave function describing an alpha 
cluster coupled to the residual daughter nucleus. 
However, such a microscopic undertaking is still considered to be a difficult task 
\cite{nirenreview} 
and the general approach is to determine $P_{\alpha}$ simply as a ratio, 
\begin{equation}\label{preformprob} 
P_{\alpha} \, = \, \tau_{1/2}^{theory} \displaystyle / \tau_{1/2}^{exp}  \, . 
\end{equation}
We refer the reader to the review article by Ni and Ren \cite{nirenreview}
(see section 2.5) for a detailed discussion on this subject. In 
Table \ref{tablefirst} we list the half-lives 
calculated in the present work (for the cases which will be studied later 
in the nonlocal framework) 
within the double folding model described above. 
The experimental half-lives 
\cite{halflifedata} and the corresponding 
values of $P_{\alpha}$ calculated using Eq. (\ref{preformprob}) are also listed in 
Table \ref{tablefirst}. These values are close to some others found in literature 
(we refer the reader once again to \cite{nirenreview} for the several references 
listing these values using different models for alpha decay). As an example, we 
mention here a microscopic calculation of the alpha cluster 
preformation probability and the decay width, presented for $^{212}$Po, within a 
quartetting function approach \cite{XuRenPRC93,XuRenPRC95}. Considering 
the interaction of the quartet with the core nucleus, $^{208}$Pb, within the 
local density approximation, the authors obtain $P_{\alpha}$ = 0.367 and 0.142 
\cite{XuRenPRC93}
using two different models for the core nucleus. In \cite{XuRenPRC95}, the calculations 
were extended to evaluate $P_{\alpha}$ for several isotopes of Po. It is gratifying 
to note that the values in Table \ref{tablefirst} for $^{210}$Po and 
$^{212}$Po are close to those found by the microscopic calculations in 
\cite{XuRenPRC93,XuRenPRC95}. 

\begin{table}[t]
 \caption{Comparison of the alpha decay half-lives evaluated using the 
double folding model with experiment \cite{halflifedata}. 
The last column lists 
the cluster preformation probability, $P_{\alpha}$, given by Eq.(\ref{preformprob}).  
}\label{tablefirst}
\begin{ruledtabular}
\centering
\begin{threeparttable}
\begin{tabular}{|c|c|c|c|p{0.12\linewidth}| }
 &Q-Value & $\tau_{1/2}^{exp}$ & $\tau_{1/2}^{theory}$ & $P_{\alpha}$  \\
 &[MeV]   & [s]                & [s]                   &   \\
\hline
 $^{254}$Fm   & 7.307 & 1.2 $\times$ 10$^4$   & 0.9 $\times$ 10$^4$ & 0.75  \\
 $^{212}$Po   &8.954  &  2.99$\times$10$^{-7}$ & 6.48$\times$10$^{-8}$ & 0.22  \\
 $^{210}$Po  & 5.407 & 1.2 $\times$ 10$^7$  &4.2 $\times$ 10$^5$    & 
0.035\tnote{a} 
\\
 $^{180}$W   & 2.515 & 5.7 $\times$ 10$^{25}$   &  1.2 $\times$ 10$^{25}$  & 0.21   \\
 $^{168}$Pt  & 6.989 & 2 $\times$ 10$^{-3}$  & 0.68 $\times$ 10$^{-3}$   & 0.34   \\
 $^{144}$Nd  & 1.903 & 7.1 $\times$ 10$^{22}$  & 5.1 $\times$ 10$^{22}$   & 0.72  \\
 $^{106}$Te  & 4.290 & 7 $\times$ 10$^{-5}$   & 2.4 $\times$ 10$^{-5}$    & 0.34   \\
\end{tabular}
\begin{tablenotes}
\item[a] The small value of $P_{\alpha}$ can be attributed to
the magic number of neutrons, N = 126, in  $^{210}$Po (see Fig. 2(c) in
\cite{weMagic} and the corresponding text for a detailed discussion).
\end{tablenotes}
\end{threeparttable}
\end{ruledtabular}
 \end{table}

Finally, we must mention that the objective of the present work is not to evaluate the 
exact half lives but rather compare the effects of nonlocalities in different models. 
Hence, we shall set $P_{\alpha}$ = 1 when we compare the 
half-lives calculated within the different models for nonlocality. 

\section{Nonlocal nuclear potentials and their local equivalent forms}\label{nonlocalpots}
The general form of the Schr\"odinger equation in the presence of nonlocality can 
be written as, 
\begin{equation}\label{nonlocaleq1}
-{\hbar^2 \over 2 \mu} \nabla^2 \Psi({\bf r}) + [V_L({\bf r}) - E] 
\Psi({\bf r}) 
= - \int \, d{\bf r}^{\prime}\, 
V_{NL}({\bf r}, {\bf r}^{\prime} ) \Psi({\bf r}^{\prime})\, , 
\end{equation}
where $V_L$ can be some isolated local potential and $V_{NL}$ the nonlocal one. 
The sources of nonlocalities in literature are globally classified into two types: 
the Feshbach and the Pauli nonlocality \cite{balantekin}. 
The Feshbach nonlocality is attributed 
to inelastic intermediate transitions in scattering processes. 
In other words, 
the description of 
an excitation at a point {\bf r} in space followed by an intermediate state 
which propagates and de-excites at some point {\bf r}$^{\prime}$ to get back to the 
elastic channel is contained in the right hand side of Eq. (\ref{nonlocaleq1}). 
Such a coupling gives rise to a coupled channels 
Schr\"odinger equation which can in principle be quite difficult to handle.

The Pauli nonlocality is attributed to the exchange effects which require 
antisymmetrization of the wave function between the projectile and the target. 
This kind of nonlocality is usually described in literature 
\cite{pereybuck,chamon1,neelam1} in terms of a 
factorized form of the potential,
\begin{equation}\label{pereybuck1}
V_{NL}({\bf r}, {\bf r}^{\prime}) = U_N\biggl ({1\over 2}|{\bf r}+{\bf r}^{\prime}|\biggr )\, 
{\exp{\biggl[ -\biggl( {{\bf r} - {\bf r}^{\prime} \over \beta}\biggr)^2  \biggr]}\over \pi^{3/2} \beta^3}\, , 
\end{equation}
involving a nonlocality range parameter $\beta$, which, in the limit $\beta \to 0$ 
brings us back to the local potential.

In what follows, we shall consider three 
different approaches to construct the effective potential ($U_L$) 
in literature which are based on this kind of description
and eventually study the manifestation of the nonlocality
in the alpha decay 
of some heavy nuclei within these models. 
Without getting into the complete details of the formalisms, 
we shall describe the three models briefly alongwith the behaviour
of the obtained effective potentials in the subsections below. 

\subsection{Perey and Buck model}\label{pereybuck}
An energy independent nonlocal potential, $U_N$, for the elastic scattering of 
neutrons from nuclei was suggested in Ref.  
\cite{pereybuck} in order to study how far the energy dependence of the phenomenological local potentials which had been used earlier could 
be accounted for by the nonlocality. The point of view taken was that though part of the energy dependence of the potentials was intrinsic, part of it came from nonlocality.  
To facilitate the numerical calculation (which involved solving the wave equation in its integro-differential form to reproduce the experimental 
data on neutron scattering up to 24 MeV), it was assumed that 
$V_{NL}({\bf r}, {\bf r}^{\prime})$ can be factorized as in Eq. (\ref{pereybuck1}). 
Apart from performing numerical calculations and fitting parameters to scattering data which were very
well reproduced, the authors provided a method to evaluate the local equivalent (LE) potentials.

Let us review
the method and the approximations used in order to later examine the 
differences in the effective potentials 
of Refs. \cite{pereybuck}, \cite{chamon1} and \cite{neelam1}. 
Using the factorized 
form of Eq. (\ref{pereybuck1}) the nonlocal Schr\"odinger equation is given as, 
\begin{eqnarray}
\biggl [{\hbar^2\over 2\mu} \nabla^2 &+& E \biggr ] \, \Psi_N({\bf r})
\label{pereybuck2}
\\
\nonumber 
&=&\int U_N\biggl ({1\over 2}|{\bf r}+{\bf r}^{\prime}|\biggr )\, 
{\exp{\biggl[ -\biggl( {{\bf r} - {\bf r}^{\prime} \over \beta}\biggr)^2  \biggr]}\over \pi^{3/2} \beta^3}\,\Psi_N({\bf r}^{\prime}) \, d{\bf r}^{\prime}\, .
\end{eqnarray}
With a change of variables, ${\bf r}^{\prime} - {\bf r} = \beta {\bf s}$, 
and using the operator form of the Taylor expansion,  the integral 
on the right hand side of Eq. (\ref{pereybuck2}) can be written as, 
\begin{eqnarray}
\nonumber
I &=& \biggl \{ \int \exp{\biggl [\beta {\bf s}\cdot
\biggl( {1\over 2} \nabla_1 + \nabla_2\biggr) \biggr ]}   \, 
{\exp{ [-s^2}]\over \pi^{3/2}}\, d{\bf s} \, \biggr \}
\\
&&\hspace{2cm} \times U_N({\bf r})\, \Psi_N({\bf r})\, ,
\label{pereybuck3}
\end{eqnarray}
where $\nabla_1$ operates only on $U_N({\bf r})$ and $\nabla_2$ on 
$\Psi_N({\bf r})$. Treating the expression [(1/2) $\nabla_1 + \nabla_2$] 
as an algebraic quantity, the authors evaluated the integral and further neglecting the effect of the operator $\nabla_1$ (i.e., assuming the potential 
$U_N(r)$ to be approximately constant), the authors obtained the 
following equation:
\begin{eqnarray}\label{pereybuck4}
\biggl [{\hbar^2\over 2\mu} \nabla^2 + E \biggr ]  \Psi_N({\bf r})& =&
U_N  \exp{[(1/4)\beta^2 \nabla^2]} \Psi_N({\bf r})\\ \nonumber
&&\hspace{-1cm}=\,  U_N [ \Psi_N({\bf r})  + (1/4) \beta^2 \nabla^2 \Psi_N({\bf r}) + \cdots ]\, .
\end{eqnarray}
Considering now the local equation
\begin{equation}\label{pereybuck5}
\biggl [{\hbar^2\over 2\mu} \nabla^2 + E \biggr ] \, \Psi_L({\bf r}) = 
U_L^{PB}\, \Psi_L({\bf r})
\end{equation}
and further assuming, 
$\Psi_N({\bf r}) \approx \Psi_L({\bf r})$, the authors obtained 
\begin{equation}\label{pereybuck6}
\nabla^2 \Psi_N({\bf r}) = - {2 \mu \over \hbar^2} (E - U_L^{PB}) \, 
\Psi_N({\bf r})
\end{equation} 
which, when substituted in Eq. (\ref{pereybuck4}) (and truncating the 
series in Eq. (\ref{pereybuck4}) up to the second term), gives
\begin{eqnarray}\label{pereybuck7}
\nonumber
\biggl [{\hbar^2\over 2\mu} \nabla^2 + E \biggr ]  \Psi_N({\bf r})& =&
U_N \Psi_N({\bf r}) \biggl [ 1  - 
{1\over 4} \beta^2 {2\mu\over \hbar^2} (E - U_L^{PB})  \biggr ]
\\ 
&&\hspace{-2cm}\simeq\, U_N \exp{\biggl (- {\mu\beta^2\over 2\hbar^2} (E - U_L^{PB}) 
\biggr) } \Psi_N({\bf r})\, .
\end{eqnarray}
Comparing the right hand sides of Eqs (\ref{pereybuck5}) and 
(\ref{pereybuck7}) (with the assumption $\Psi_N({\bf r}) \approx \Psi_L({\bf r})$), the authors finally obtained 
\begin{equation}\label{pereybuck8}
U_L^{PB}(r)\, \exp{\biggl [ {\mu\beta^2\over 2 \hbar^2} 
(E - U_L^{PB}(r) ) \biggr ]}\, =\, U_N(r)\, .
\end{equation}
The above equation was in principle derived with the assumption that 
the potential inside the nucleus is constant and the $r$ dependent form 
above was justified a posteriori from the results obtained in the paper. 
Finally, the transcendental equation (Eq. (\ref{pereybuck8})) is solved to 
obtain $U_L^{PB}(r)$. 
Taking initially the potentials $U_N$ and $U_L^{PB}$ to be constant in order to 
derive the transcendental equation, 
$U_L^{PB}\, \exp{[ {\mu\beta^2\over 2 \hbar^2} 
(E - U_L^{PB})]}\, =\, U_N$, and then introducing the $r$ dependence to get 
(\ref{pereybuck8}) introduces an inconsistency at small $r$ which will 
be discussed in Section \ref{smalldistance}.

\subsection{S\~ao Paulo potential}\label{saopaulo}
Based on conceptually similar considerations as of the Perey and Buck model, a slightly different form of a local equivalent potential 
was derived in Ref. \cite{chamon1} and applied successfully to reproduce several different scattering data 
\cite{saopaulo1,saopaulo2,saopaulo3}.  
The authors defined the 
local equivalent potential as 
\begin{equation}\label{saopaulopot1}
U^{SP}_L(r) \approx V_n(r)\, \exp{(-\gamma\, [E - V_C(r) - U^{SP}_L(r)])}\, 
\end{equation} 
where $\gamma = \mu \beta^2 / 2 \hbar^2$ and $V_n$ is the 
double folding nuclear potential as described in Section \ref{doublefold}. 
The authors cautioned that the local equivalent potential $U_L^{SP}$ is very well described 
by the above equation except for small distances (i.e., $r \to 0$). Further identifying the factor in 
the exponential with a velocity, 
\begin{equation}
v^2 = {2 \over \mu} \,E_k(r) = {2 \over \mu}\, [E - V_C(r) - U^{SP}_L(r)]\, , 
\end{equation}
the authors mentioned that the effect of the Pauli nonlocality is equivalent to a velocity dependent 
nuclear potential.
Note that the local equivalent S\~ao Paulo potential of Eq. (\ref{saopaulopot1}) is in principle the same as that 
proposed by Perey and Buck (in Eq. (\ref{pereybuck8})) 
if we substitute $U_N$ by the double folding potential $V_n(r)$  
and neglect the Coulomb potential $V_C$ in Eq. (\ref{saopaulopot1}).

Both the local equivalent potentials, $U^{SP}_L$ and $U^{PB}_L$ 
are energy dependent and as will be 
seen later, approach a finite value as $r \to 0$. 

\subsection{Mumbai potential}\label{mumbai}
In Ref. \cite{neelam1}, 
the authors proposed a novel method to solve the integro-differential equation
in Eq. (\ref{nonlocaleq1}). The method 
which was introduced in Ref. \cite{neelam1} 
involved the use of  the mean value theorem of integral calculus to 
obtain an effective potential, 
which, in contrast to the methods discussed so far, 
was found to be energy independent. 
Apart from relying on the mathematical validity, the method was further
tested by calculating the total and differential cross sections
for neutron scattering off $^{12}$C, $^{56}$Fe and $^{100}$Mo nuclei in 
the low energy region (up to 10 MeV) and reasonably good agreement with data was found. 
We shall refer to this approach of Ref. \cite{neelam1} as the Mumbai approach 
and briefly review the main steps in their derivation below.

Performing a partial wave expansion of $V_{NL}({\bf r}, {\bf r}^{\prime})$ and $\Psi({\bf r}^{\prime})$ in Eq. (\ref{nonlocaleq1}), 
one obtains the radial equation, which, in the 
absence of the spin-orbit interaction is given as 
\begin{widetext}
\begin{eqnarray}\label{nonlocalradial}
&&{\hbar^2 \over 2 \mu}\, \biggl ( {d^2\over dr^2} \,- \, 
{l(l+1) \over r^2} \biggr ) \, u_{l}(r) + E u_{l}(r)  
\, = \, \int_0^{\infty} \, g_l(r,r^{\prime}) u_{l}(r^{\prime})\, dr^{\prime}\, ,
\\
{\rm where,} \hspace{1cm}&&
g_l(r,r^{\prime}) = \biggl ( {2 r r^{\prime} \over \sqrt{\pi} \beta^3} \biggr )\, \exp{\biggl ({-r^2 - r^{\prime\,2}\over \beta^2}\biggr )}\, 
\int_{-1}^1\, U_N\biggl({ |{\bf r} + {\bf r}^{\prime}|  \over 2}\biggr)  \, \exp{\biggl ({2 r r^{\prime} x\over \beta^2}\biggr )}\, P_l(x) \, dx \, . 
\label{gkernel}
\end{eqnarray}
\end{widetext}
Making use of the mean value theorem to rewrite the integral on the right hand side of Eq. (\ref{nonlocalradial}) and 
restricting the upper limit of integration to the range of the nuclear interaction, after some algebra, the authors 
obtain an effective potential given by \cite{neelam1}, 
 \begin{equation}\label{mumbaipot}
U_L^M(r) = \int_0^{r_m}\, g_l(r,r^{\prime})\, dr^{\prime}\, 
\end{equation}
where, $g_l(r,r^{\prime})$ is written as in Eq. (\ref{gkernel}). 
Note however that the Mumbai (M) 
potential, in contrast with that of the Perey-Buck (PB) model and the 
S\~ao Paulo (SP) potential, does not depend on energy. Indeed, it also displays a different behaviour at 
small distances with $U_L^M \to 0$ for $r \to 0$.  

\subsection{Behaviour of $U_L$ in the three models}
\begin{figure*}
\includegraphics[scale=1]{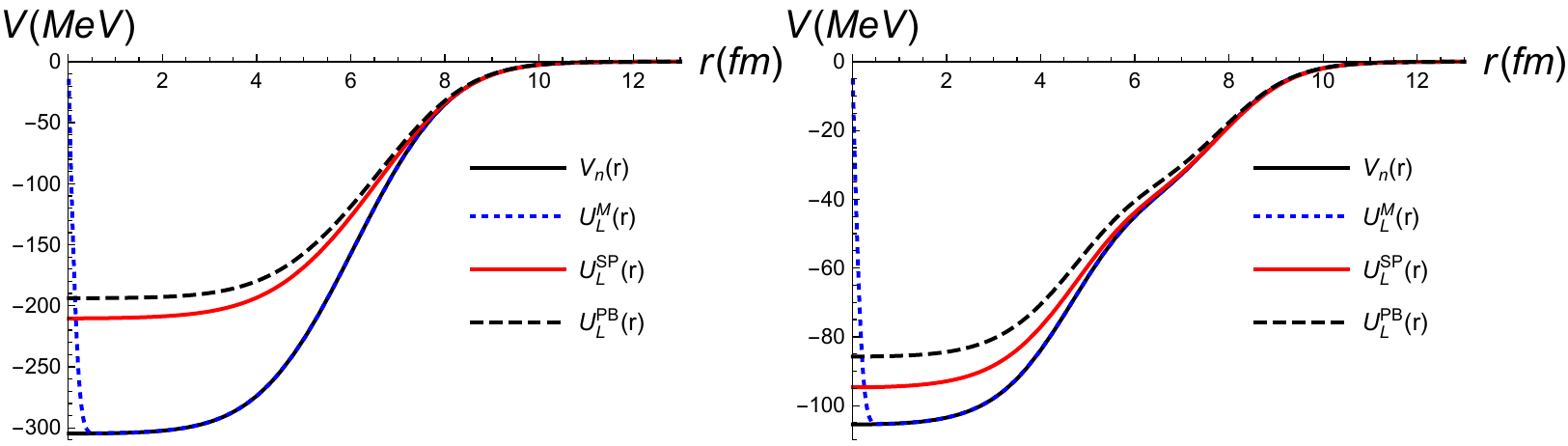}
\caption{The strong interaction potential between the alpha ($^4$He) 
and the daughter nucleus ($^{206}$Pb) in the decay of $^{210}$Po. 
The left panel shows a comparison of the potentials, 
$U_L^{SP}(r)$ (dot-dashed line), $U_L^{PB}(r)$ (dashed line), 
$U_L^M(r)$ (dotted line) and the double folding potential $V_n(r)$ (solid line) 
with the knock-on exchange term included.  
The right panel shows the same without the exchange term.
For the sake of this comparison, we choose $\lambda$ = 1 
in Eq. (\ref{potnucl}).}
\end{figure*}

In Fig. 1, we compare the 
effective potentials with the double folding potential $V_n(r)$. 
To perform this comparison, we choose $\lambda$ = 1 in Eq. (\ref{potnucl}).
Thus, replacing $U_N(r)$ with $V_n(r)$ of Eq. (\ref{potnucl}), the Perey-Buck 
local equivalent potential ($U_L^{PB}(r)$) is 
evaluated using Eq. (\ref{pereybuck8}),    
the S\~ao Paulo local equivalent 
potential ($U_L^{SP}(r)$) is evaluated using (\ref{saopaulopot1}) 
and the local effective Mumbai potential ($U_L^M(r)$)
using Eq. (\ref{mumbaipot}). 
Since the exchange term in 
Eq. (\ref{nnfree}) is often not included in the 
calculation of the 
local equivalent potentials \cite{neelam1} (in order to avoid double counting of the Pauli nonlocality), we show the 
potentials with (left panel) as well as without (right panel) 
this term included. The three potentials, $U_L^{PB}(r)$, $U_L^{SP}(r)$ and 
$U_L^M(r)$ are evaluated for a double folding potential between an alpha and $^{206}$Pb nucleus, which 
are the decay products (and hence originally the cluster nuclei) in the alpha decay of $^{210}$Po. 
The nonlocality parameter $\beta$ is taken to be 0.22 fm 
(explanation given in section \ref{results}) , 
following the prescription given in \cite{jackson}. 
The S\~ao Paulo and Perey-Buck potentials are quite similar as expected while the energy independent 
Mumbai potential as mentioned earlier behaves differently at small $r$. 
The latter, as we shall see in the 
next section with the example of a simple form for $U_N$, 
follows quite simply from the nonlocal radial equation. 
The discrepancy between $U_L^{PB}$, $U^{SP}_L$ and $U^M_L$ 
probably arises due to the 
assumptions in the derivation of Eq. (\ref{pereybuck8}).  

\section{Model dependence at small distances}\label{smalldistance}
With the aim of understanding the difference in the small $r$ 
behaviour of the effective 
potentials mentioned in the previous section, we shall now 
analyze the nonlocal kernel using a simple rectangular well for the nuclear 
potential and try to obtain analytical expressions.
Let us begin by considering the integral on the  
right hand side of Eq. (\ref{nonlocalradial}), namely,  
$\displaystyle{\int_0^{\infty} \, g_l(r,r^{\, \prime})\, u_l(r, r^{\, \prime}) \, dr^{\,\prime}}$. 
Given the fact that $g_l(r,r^{\,\prime})$ is peaked close to $r=r^{\, \prime}$ 
(see for example Figs 1a and 5a in Ref. \cite{neelam1}), we
perform a Taylor expansion of the wave function about $r = r^{\, \prime}$ and  
write the above integral as 
\begin{equation}\label{expansion}
u(r) \, \int_0^{\infty} \, g_l(r,r^{\, \prime}) dr^{\, \prime} \, +\, 
\int_0^{\infty} \, (r^{\,\prime} - r)\,u^{\,\prime}(r)\, g_l(r,r^{\, \prime}) 
dr^{\, \prime}\, +\, \cdots
\end{equation} 
For the case of a rectangular well of depth -$U_0$ and range $R$, i.e., 
for $U_N(r) = - U_0 \Theta(R - r)$ (where $\Theta$ is the Heaviside step function) 
and assuming $l = 0$ for simplicity, 
we can evaluate the first integral in (\ref{expansion}) analytically. 
Retaining only the first term in the expansion (\ref{expansion}) we can define,
\begin{equation}\label{newlocal}
U_L(r) = \int_0^{\infty} \, g_l(r,r^{\, \prime}) dr^{\, \prime} 
\end{equation}
with, 
\begin{widetext} 
\begin{eqnarray}\label{gkernel2}
g_l(r,r^{\, \prime}) &=& 
{2 \over \sqrt{\pi} \beta} \, 
U_N\biggl[ {1\over 2} (r + r^{\prime}) \biggr ] \, 
\exp{\biggl [ - {(r^2 + r^{\prime\,2})\over \beta^2} 
\biggr ]} \sinh{{2 r r^{\prime} \over \beta^2}} \nonumber \\
&=& -
{2 \over \sqrt{\pi} \beta} \, 
U_0 \, \Theta \biggl(R - {r + r^{\prime} \over 2} \biggr ) \, 
\exp{\biggl [ - {(r^2 + r^{\prime\,2})\over \beta^2}
\biggr ] }\, 
\sinh{{2 r r^{\prime} \over \beta^2}} \, . 
\end{eqnarray}
\end{widetext}
One can see that $\Theta$ = 1 only for $2R - r - r^{\prime} >$ 0, 
i.e., $2R - r > r^{\prime}$ and the upper limit of integration 
in (\ref{newlocal}) changes from $\infty$ 
to $2R - r$. If $r > 2R$, $r^{\prime}$ is negative. Hence, to 
ensure that $r \le 2R$ we write,
\begin{widetext} 
\begin{eqnarray}\label{proof1}
U_L(r) &=&
{-2 U_0 \over \sqrt{\pi} \beta} \, \Theta(2R - r)\, 
\int_0^{2R -r} \, dr^{\, \prime} 
\, \exp{\left[-\,{(r^2+r^{\,\prime\, 2}) \over \beta^2}\right]}\, \sinh{{2rr^{\,\prime} \over \beta^2}}
\\\nonumber 
&=&  - {U_0\over 2}\, \Theta(2R -r)\, 
\biggl[ \, erf\biggl ( {2(R-r)\over \beta}\biggr ) + 
2 erf\biggl({r\over \beta} \biggr ) - erf\biggl ({2R\over \beta}\biggr )\, \biggr ].
\end{eqnarray}
In the limit, $r \to 0$ (for $\beta > 0$), 
since $erf(0) = 0$, the potential $U_L(r)$ vanishes. 
When $r$ is finite, we must consider two cases: 
\begin{eqnarray}\label{proof1p}
U_L(r) &=& - {U_0\over 2}\, 
\biggl[ \, erf\biggl ( {2(R-r)\over \beta}\biggr ) + 
2 erf\biggl({r\over \beta} \biggr ) - erf\biggl ({2R\over \beta}\biggr )\, \biggr ] 
\, \, \, \, \,\,\,\,\,\,\,\,\forall \, \,r < R \\
&=& - {U_0\over 2}\, 
\biggl[ \, - erf\biggl ( {2|R-r|\over \beta}\biggr ) + 
2 erf\biggl({r\over \beta} \biggr ) - erf\biggl ({2R\over \beta}\biggr )\, \biggr ] 
\, \, \, \, \,\,\,\,\forall \, \,R < r < 2R \, .
\end{eqnarray}
\end{widetext}
If $\beta \to$ 0 (i.e. in the absence of nonlocality), 
since $erf(\infty) = 1$, $U_L(r) = - U_0$ for 
$r < R$ and 0 for $R < r < 2R$, as expected.


The above derivation on the one hand justifies the behaviour of the 
Mumbai potential at small $r$ but 
on the other hand displays an inconsistency between Eqs (\ref{proof1}) and 
(\ref{pereybuck8}). 
$U_L^{PB}(r)$ in Eq. (\ref{pereybuck8}), approaches a finite value as $r \to 0$ 
(for finite $\beta$) as we already noticed in 
Fig. 1 with a more realistic form of $U_N(r)$. However, $U_L(r)$, 
as derived above from the radial nonlocal equation vanishes for $r \to 0$.   
 
Since the starting point for 
the derivation of the Mumbai potential is indeed the nonlocal radial equation, 
it seems to be in agreement with the 
behaviour of $U_L(r)$ as in Eq.(\ref{proof1}) but not with that in 
Eq. (\ref{pereybuck8}). 
The inconsistency between Eq. (\ref{pereybuck8}) and Eq. (\ref{proof1}) probably arises due 
to the approximations made in the derivation of Eq. (\ref{pereybuck8}). 

\section{Iterative schemes}\label{iterations}
Models for the nonlocal nuclear interaction are usually tested for their validity by 
reproducing scattering data. The solution of the radial equation 
(\ref{nonlocalradial}) is obtained by implementing an iterative procedure. The starting 
point of the iterative procedure involves an effective or 
local equivalent potential which is a solution of the homogeneous equation such 
as Eq. (\ref{pereybuck5}). For example, the iteration scheme can be started with the 
local equation, 
\begin{equation}\label{localradial}
{\hbar^2 \over 2 M}\, \biggl ( {d^2\over dr^2} -  
{l(l+1) \over r^2} \biggr )  u^0_{l}(r) + (E - U_L(r)) u^0_{l}(r)  
 = 0
\end{equation}
and followed by 
\begin{eqnarray}
\label{ieratenonlocal}
&& {\hbar^2 \over 2 M}\, \biggl ( {d^2\over dr^2} -  
{l(l+1) \over r^2} \biggr )  u^i_{l}(r) + (E - U_L(r)) u^i_{l}(r)  
\\
\nonumber
&&\hspace{1cm} = \, \biggl [\int_0^{R} \, g_l(r,r^{\prime}) u^{i-1}_{l}(r^{\prime})\, dr^{\prime}\,
-\, U_L(r) u^{i-1}_{l}(r) \biggr]\,,
\end{eqnarray}
where the suffix $i$ denotes the $i^{th}$ order approximation to the correct
solution. The upper limit $R$ is the radius at which the contribution of the 
kernel becomes negligible. 
The iteration is continued until the logarithmic derivative at $R$ obtained from 
$u^i_{l}(r)$ agrees up to a certain reasonable precision with the one calculated from 
$u^{i-1}_{l}(r)$. Generally one finds that a few iterations 
\cite{pereybuck,neelam1,neelam2} already lead to a good agreement with data. 

The effect of the nonlocal potentials could in principle be tested by calculating 
the half lives of radioactive nuclei. Restricting ourselves to the 
discussion of alpha decay, one could follow a similar iteration scheme as above, 
however with the difference that $u_l(r)$ would have different boundary conditions. 
Considering the decaying nucleus as a resonant state (and noting that there 
are no incident particles), the solution of the radial 
equation would be a ``Gamow function" \cite{mondragon} which vanishes at the origin 
and behaves as a purely outgoing wave asymptotically. The so-called correct solution 
obtained from such an iterative scheme could then be used in a quantum mechanical 
description of the alpha particle decay rates.
Such an analysis of the alpha decay of several nuclei could serve as a complementary 
tool for fixing the parameters or assumptions of the nonlocal models.
In order to find out if such a task is worth undertaking, in the 
present work we take the first step of comparing the 
alpha decay half lives of some heavy nuclei using 
different models of the local equivalent (or effective) potentials which satisfy the homogeneous 
equation. The latter allows us to follow the procedures outlined in Section 
\ref{alphadecay} to evaluate the half life within the JWKB approximation where the 
wave function is a solution of the homogeneous equation. 

\section{Results and Discussions}\label{results}
To study the effect of nonlocality in alpha decay, we evaluate the half-lives 
of some spherical nuclei (with spin-parity, $J^P$ = 0$^+$), decaying in 
the $s$-wave.  
In order to calculate the half-lives, we use the density dependent 
double folding model 
introduced in Section \ref{doublefold} as input for the evaluation of the 
effective potentials, $U_L$. 
Note that the nonlocality appears only in the strong part of the potential to which we add the 
Coulomb and the centrifugal part as given in Section \ref{doublefold}. 
Since the half-lives 
are evaluated within the semi-classical JWKB approximation, the potentials are required to 
satisfy the Bohr-Sommerfeld condition in Eq. (\ref{bohrsomer}) which then fixes the strength of 
$\lambda$ in Eq. (\ref{potnucl}). 
In Fig. 2, we compare the full  
potentials (i.e., Eq. (\ref{pot1}) for the double folding potential and 
$U_L(r)\,+\,V_C(r)\,+\,{\hbar^2\,(l\,+\,1/2)^2 \over \mu\,r^2}$ 
for the PB, SP and M cases) 
evaluated for the interaction between $^4$He ($\alpha$) and $^{206}$Pb which form 
the cluster in $^{210}$Po.  
In order to evaluate the energy dependent Perey-Buck and S\~ao Paulo potentials, 
we assume the energy to be the $Q$-value 
(which is approximately the kinetic energy of the 
alpha in the final state) in the decay of $^{210}$Po.  
\begin{figure*}[ht]
\includegraphics[scale=1]{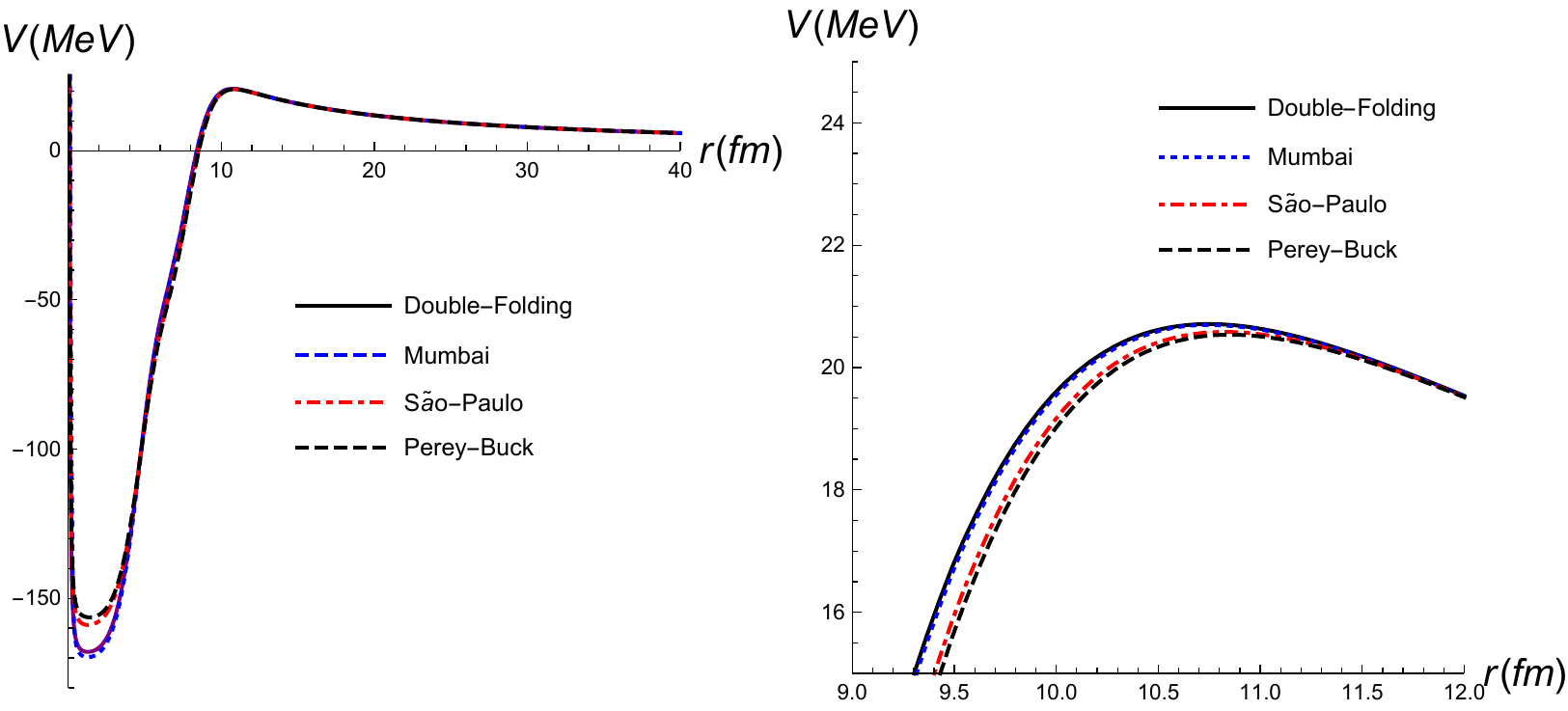}
\caption{Comparison of the local equivalent potentials for the interaction between $^4$He and $^{206}$Pb.
The full potentials (including the strong and Coulomb interaction as well as the centrifugal part arising due to 
the Langer term) are displayed in the left panel. The right panel shows the Coulomb barrier on a different 
scale. The knock-on exchange term is not included. The values of $\lambda$ 
appearing in Eq. (\ref{potnucl}) are listed in Table \ref{tablelast}. 
}\label{Figure2}
\end{figure*}

Using the three different models for the effective strong interaction, we now evaluate the half-lives 
in the alpha decay of medium and heavy nuclei. 
The nonlocality parameter, $\beta$ is given by $\beta = b_0 m_0 /\mu$, 
where $b_0$ is the nonlocal range of the nucleon nucleus interaction, 
$m_0$ is the nucleon mass and $\mu$ the alpha-nucleus reduced mass (see \cite{jackson} 
for details). We choose $b_0$ = 0.85 fm as in Ref. \cite{pereybuck}, so that 
$\beta$ = 0.22 fm. 
The half-lives evaluated with the nonlocality included are found to 
decrease as compared to those evaluated using the double folding model 
without nonlocality.

The percentage decrease in half-life due to nonlocality is defined as, 
\begin{equation}\label{percentage}
PD = {\tau_{1/2}^{DF} - \tau_{1/2}^{NL} \over \tau_{1/2}^{DF} }\, \times\, 100\, ,
\end{equation} 
where, $\tau_{1/2}^{DF}$ is the half-life evaluated in the double folding model 
without the inclusion of nonlocalities and $\tau_{1/2}^{NL}$ is the one evaluated using 
the different nonlocal frameworks. We remind the reader that we choose 
the cluster preformation factor $P_{\alpha}$ = 1 for this comparison. 

In Table \ref{table1} we list the percentage decrease in the half-lives of 
several nuclei. The numbers outside parentheses correspond to 
calculations with the so called knock-on exchange term 
excluded in order to avoid double counting of the Pauli nonlocality. Inclusion 
of this term (numbers in parentheses) causes a larger percentage decrease of 
half-lives due to nonlocality within the Perey-Buck and S\~ao Paulo models. 
The effect of nonlocality is in general small within the Mumbai approach. 
Apart from this, we note that the effect of nonlocality is smaller in 
the decay of $^{212}$Po (Q = 8.954 MeV) 
as compared to $^{210}$Po (5.407 MeV). Though the difference is not large, 
it hints towards a decrease in the effect of nonlocality with 
increasing energy in the two isotopes. 

\begin{table*}[ht]
 \caption{Percentage decrease, $PD$ as in Eq. (\ref{percentage}), 
in alpha decay half-lives of nuclei using three 
different models of nonlocality. 
Numbers within parentheses include the effect of the 
so called knock-on exchange term and those outside ignore this term.
The nonlocality parameter $\beta$ = 0.22 fm.}\label{table1}
\begin{ruledtabular}
\begin{tabular}{ |p{3cm}|p{3cm}|p{3cm}|p{3cm}| }
 & Mumbai &S\~ao Paulo&Perey-Buck\\
 \hline

 $^{254}$Fm  & 2.49 (3.85)  &  31.7 (69.8)  &  40.3 (71.5) \\
 $^{210}$Po  & 3.4 (2.4)  &  29.9 (66.2)  &  37.3 (67.9) \\
 $^{212}$Po  & 3.26 (3.28) &  26.7(62.0)   & 33.0 (63.8) \\
$^{180}$W   & 3.8 (4.0) &  30.1 (66.1) &  37.6 (67.7) \\
 $^{168}$Pt  & 4.9 (4.1) &  27.5 (62.7)  &  34.9 (64.9)\\
 $^{144}$Nd  & 3.9 (3.8) &  25.9 (60.3)  &  31.2 (61.9)\\
 $^{106}$Te  & 3.2 (4.1) &  19.4 (51.9) &  24.1  (53.6)\\

\end{tabular}
\end{ruledtabular}
\end{table*}

In order to understand the results in Table \ref{table1}, let us examine 
Fig. \ref{Figure2} and the factors presented in Table \ref{tablelast}. 
We consider the numbers outside the parentheses in Table \ref{table1} which 
means that the knock-on exchange term is not included in the calculation of $V_n$. 
The full potentials given in Fig. \ref{Figure2} are obtained using 
Eqs (\ref{pot1}) and (\ref{potnucl}) with the values of $\lambda$ 
(listed in Table \ref{tablelast}) being fixed 
by the Bohr-Sommerfeld condition (\ref{bohrsomer}). 
Assuming $P_{\alpha}$ = 1 and rewriting the expression for the 
width in (\ref{gurwidth}) as 
$$\Gamma ={\hbar^2 \over 2 \,\mu}\ N P\, ,$$
where, 
the normalization factor 
$N =[\,\int_{r_1}^{r_2}\,
{dr /k(r)}\,]^{-1} $
and the exponential factor or the penetration probability, 
$P = e^{-2\int_{r_2}^{r_3}\,k(r)\,dr}$, 
we note from Table \ref{tablelast} that it is indeed the difference in the 
penentration probabilities, $P$, which leads to the differences in the 
percentage decrease in the half-lives in Table \ref{table1}. 
The normalization factors are almost constant in all models. This fact is also 
reflected in Fig. \ref{Figure2}. In the right panel we notice that the 
Coulomb potential in the S\~ao Paulo and Perey-Buck models is shifted to the right 
as compared to the double-folding and Mumbai potentials. This shift leads 
to a shift of the second turning point, $r_2$, to bigger values and hence smaller
half-lives 
(due to the bigger exponential factor as can be seen in Table \ref{tablelast}). 

\begin{table*}[ht]
\caption{Factors contributing to the calculated half-lives in different models. 
The strength of the strong interaction, $\lambda$, which is fixed by the Bohr-Sommerfeld 
condition (\ref{bohrsomer}), the normalization factor 
N = $[\int_{r_1}^{r_2} dr/k(r)]^{-1}$ and the penetration 
probability, P$=e^{-2\int_{r_2}^{r_3}\,k(r)\,dr}$, are 
listed in all models for each of the nuclei considered. The knock-on exchange term is 
not included. 
}\label{tablelast}
\begin{tabular}{ |c|c||c|c|c||c|c|c||c|c|c||c|c|c|}
\hline
{\multirow{4}{*}{Isotope}} & {\multirow{4}{*}{Q-Value}} & \multicolumn{3}{c||}{Double Folding} & \multicolumn{3}{c||}{Mumbai } & 
\multicolumn{3}{c||}{S\~{a}o Paulo}& \multicolumn{3}{c|}{Perey Buck}\\
\cline{3-14}
& & & & & & & & & & & & &\\
& & $\lambda$&P  &N &$\lambda$ & P & N & $\lambda$& P & N & $\lambda$& P & N \\
&[MeV] & & & [fm$^{-2}$] & &  & [fm$^{-2}$] & &  & [fm$^{-2}$] & &  & [fm$^{-2}$] \\
 \hline

 & & & &  & & & & & & & & & \\
 $^{254}$Fm & 7.307 &1.95 & $3\times10^{-25}$&0.34  &1.96 &$3.1\times 10^{-25}$&0.34  
&2.08 & $4.5\times 10^{-25}$ &0.34 &2.30 &  $5.2\times 10^{-25}$&0.34\\
 $^{212}$Po & 8.954 &2.02 &$3\times 10^{-14}$&0.34 &2.03 &$3.2\times 10^{-14}$&0.34&2.17 &
$4.2\times 10^{-14}$&0.33  &2.36 & $4.6\times 10^{-14}$&0.33 \\
 $^{210}$Po &5.407 &2.10 & $5.4\times 10^{-27}$&0.36 &2.12 &$5.6\times 10^{-27}$&0.35
&2.25 & $7.8\times 10^{-27}$&0.35&2.45 &  $8.6\times 10^{-27}$&0.36 \\
 $^{180}$W  & 2.515 &2.04& $2\times 10^{-46}$&0.36 &2.05 & $2.1\times 10^{-46}$&0.36
&2.17 &$2.8\times 10^{-46}$&0.36 &2.35& $3.1\times 10^{-46}$&0.35  \\
 $^{168}$Pt &6.989 &2.08&$3\times 10^{-18}$&0.35  &2.10 & $3.2\times 10^{-18}$&0.35&2.22&
$4.2\times 10^{-18}$&0.35&2.41& $4.6\times 10^{-18}$&0.35\\
 $^{144}$Nd &1.903 &2.23&$3.6\times 10^{-44}$&0.37&2.26& $3.8\times 10^{-44}$&0.37&2.39&
$4.9\times 10^{-44}$&0.36&2.55&  $5.3\times 10^{-44}$&0.36 \\
 $^{106}$Te &4.290 &2.28 &$6.2\times 10^{-17}$&0.35 &2.30& $6.5\times 10^{-17}$& 0.35
&2.43 &$7.9\times 10^{-17}$&0.35 &2.58& $8.4\times 10^{-17}$&0.35 \\

 \hline
\end{tabular}
\end{table*}

Before closing the discussions, some comments about one of the earliest investigations 
of nonlocalities in alpha decay are in order here. In a series of works
\cite{chaudhury1,chaudhury2,chaudhury3,chaudhury4}, 
M. L. Chaudhury studied the effects of nonlocalities in alpha decay of different 
nuclei. Using an integro-differential equation similar to that of Frahn and Lemmer 
\cite{frahnlemmer}, in Ref. \cite{chaudhury1}, the author calculated 
the transmission coefficient (or penetration factor) for alpha tunneling within the 
WKB approximation. 
The point-like Coulomb field was 
superimposed by a nonlocal alpha-nucleus interaction based on the Igo 
potential \cite{igo} and a Gaussian form was used for the nonlocal function with a nonlocality 
parameter of $\beta$ = 0.9 fm. Investigating for the alpha decay of 
$^{254}$Fm in Ref. \cite{chaudhury1}, the 
penetration factor was found to increase by a factor of 1.7 due to nonlocality. 
This amounts to a decrease of about 40 \% in the half-life. 
The investigation in Ref. \cite{chaudhury1} was  
extended to several even-even nuclei and the author once again found a large 
increase ($\sim$ 50 \%) \cite{chaudhury2} 
in the penetration factors due to nonlocality.
The author further studied the effects of nonlocality 
in deformed and rare earth nuclei \cite{chaudhury3,chaudhury4} with the 
inclusion of an exchange term in the nonlocal kernel. 
Though the calculation in Ref. \cite{chaudhury1} involves a different 
nonlocal framework as compared to the models considered in the present work, 
the 40\% decrease 
in the value of the half-life of $^{254}$Fm \cite{chaudhury1} 
(evaluated without the exchange term in the nonlocal kernel), 
is similar to the Perey-Buck model result in Table \ref{table1},  
without the inclusion of the knock-on exchange term. 

\section{Summary}\label{conclude}
Investigation of the effects of nonlocality in the nuclear interaction 
began several decades ago but has remained to be a topic of continued interest until now. 
The vast majority of works proposing different approaches to understand the 
origin as well as the manifestations of the nonlocality concentrate on the reproduction
of scattering data. Here, we have proposed the study of the effects of nonlocality 
on alpha decay half lives of nuclei as a complementary tool for determining the 
nonlocal interaction within different models. 

To be specific, we have studied these effects using three different models available 
in literature. Though all the three models agree qualitatively on the result that 
the nonlocal nuclear interaction leads to a decrease in half-lives, the percentage 
decrease in the three models is quite different. The recent Mumbai model 
\cite{neelam1,neelam2} predicts a very small percentage decrease of 
about 2 -  5\% in 
most heavy nuclei studied, 
however, the Perey-Buck and S\~ao Paulo models predict a much 
bigger decrease of around 20 - 40\% 
(in the absence of the knock-on exchange term). 

In order to understand the above differences, we examined 
the model assumptions in detail and found an inconsistency in the behaviour of the 
local equivalent potentials, $U_L(r)$, derived by starting with 
the 3-dimensional Schr\"odinger equation and the radial one. 
Whereas the former leads to 
$U_L(r)$ which is finite at the origin, the latter vanishes for $r \to 0$. 
Indeed the Perey-Buck and the S\~ao Paulo models are of the former type 
(with the local equivalent potential being  
finite at $r$ = 0) but the effective potential of the Mumbai group 
vanishes at $r$ = 0. Introducing the nonlocal framework 
to find the Gamow functions corresponding to the decaying nuclei 
and performing a more exact quantum mechanical calculation of the half lives 
to compare with data could possibly provide a better explanation of the different 
behaviours of the potentials.

\begin{acknowledgments}
The authors are thankful to B. K. Jain for many useful discussions.
N.G.K. thanks the Faculty of Science, Universidad de los Andes, Colombia for 
financial support through grant no. P18.160322.001-17. N.J.U. acknowledges 
financial support from SERB, Govt. of India (grant number YSS/2015/000900).
\end{acknowledgments}

\end{document}